\documentclass[conference]{IEEEtran}
\usepackage{textcomp}
\usepackage{cite}
\usepackage{subfigure}

\ifCLASSINFOpdf
   \usepackage[pdftex]{graphicx}
\else
\fi
\usepackage{amsmath}
\usepackage{algorithmic}
\usepackage{url}

\usepackage{RobStd}

\begin{document}

\HideNotes

\title{Application-level Studies of Cellular Neural Network-based Hardware Accelerators}

\author{Qiuwen Lou$^{1}$, Indranil Palit$^{1}$, Andras Horvath$^{2}$, Li Tang$^{3}$, Michael Niemier$^{1}$, X. Sharon Hu$^{2}$\\ 
$^{1}$Department of Computer Science and Engineering, University of Notre Dame\\
$^{2}$Faculty of Information Technology, Pazmany Peter Catholic University\\
$^{3}$Los Alamos National Laboratory\\
\textit{\{qlou, ipalit, mniemier, shu\}@nd.edu, horvath.andras@itk.ppke.hu, ltang@lanl.gov}
\vspace{-16pt}
}

\maketitle

\begin{abstract}
As cost and performance benefits associated with Moore\textquotesingle s Law scaling slow, researchers are studying alternative architectures (e.g., based on analog and/or spiking circuits) and/or computational models (e.g., convolutional and recurrent neural networks) to perform application-level tasks faster, more energy efficiently, and/or more accurately.  We investigate cellular neural network (CeNN)-based co-processors at the application-level for these metrics. While it is well-known that CeNNs can be well-suited for spatio-temporal information processing, few (if any) studies have quantified the energy/delay/accuracy of a CeNN-friendly algorithm and compared the CeNN-based approach to the best von Neumann algorithm at the application level.  We present an evaluation framework for such studies.  As a case study, a CeNN-friendly target-tracking algorithm was developed and mapped to an array architecture developed in conjunction with the algorithm.  
We compare the energy, delay, and accuracy of our architecture/algorithm (assuming all overheads) to the most accurate von Neumann algorithm (Struck).  Von Neumann CPU data is measured on an Intel i5 chip.  The CeNN approach is capable of matching the accuracy of Struck, and can offer approximately {$\mathbf{1000 \times}$} improvements in energy-delay product.  
\end{abstract}

\IEEEpeerreviewmaketitle

\section{Introduction}
As Moore's Law-based device scaling/accompanying performance scaling trends slow, there is continued interest in new technologies and computational models to perform a given task faster and more energy efficiently \cite{2015_Nikonov, 7533513}.
Furthermore, there is growing evidence that, with respect to traditional Boolean circuits, it will be challenging for emerging devices to compete with CMOS technology.  \cite{2015_Nikonov, 7533513, Benchmarking_Center}.  

Researchers are increasingly looking for new ways to exploit the unique characteristics of emerging devices (e.g., non-volatility) as well as architectural paradigms that transcend the energy efficiency wall.  In this regard, cellular neural networks (CeNNs) are now under investigation via the Semiconductor Research Corporation's benchmarking activities \cite{Benchmarking_Center, STARnet} as {\bf (i)} they can solve a broad set of problems \cite{CeNN} (e.g,. image processing, associative memories, etc.), and {\bf (ii)} can exploit the unique properties of both spin- and charge- based devices \cite{7533513, ms-cnn}.  However, we must consider what application spaces/computational models might benefit from CeNNs, and if CeNNs outperform other alternative architectures/models, for the same problem, at the application-level.

We have developed a framework for evaluating CeNN co-processors to quantitatively assess CeNNs at the application-level.  {\bf (i)} We begin with algorithm development where processing tasks are mapped to CeNNs or more conventional CPUs/GPUs (e.g., for image recognition, CeNNs can be highly efficient for feature extraction tasks given the architecture's parallel nature; for more mathematical operations, CPUs may be more efficient.)  {\bf (ii)} Next, given the analog nature of a CeNN, and the inherent nature of inference applications, algorithmic accuracy must be evaluated at multiple levels (e.g., we must address overall algorithmic quality, and any impact on algorithmic quality due to lower precision hardware.)  Algorithmic refinement may be needed.  {\bf (iii)} Algorithms must then be mapped to a suitable hardware architecture (e.g., parallel CeNNs vs. a CeNN that is used serially).  Again, algorithm refinement may be necessary.  {\bf (iv)} Finally, we must compare energy, delay, {\it and accuracy} projections to the best von Neumann algorithm for the same application-level problem.

As a case study, we have developed a CeNN-based target tracking algorithm to complete an {\it application-level case study}.  Target tracking is a fundamental component of many computer vision applications \cite{Wu13}.
We have also developed a CeNN array architecture for our algorithm with analog CeNN circuitry, analog-to-digital (ADC) and digital-to-analog (DAC) converters, and an 8-bit register.  We compare our approach to the best reported von Neumann algorithm in terms of energy, delay, and accuracy.  Per \cite{Wu13}, we have identified the best von Neumann algorithm (Struck \cite{Hare11}) for images sequences in the VisualTracking.net benchmark suite \cite{benchmark}. 
Our CeNN-based approach (with ADC/DAC overheads, etc.) is competitive with Struck in terms of accuracy, and improvements in energy-delay products (EDP) can be $>$1000X.  Finally, we show how our algorithm can be tuned such that a CeNN approach becomes competitive for sequences where Struck has superior accuracy.  

\section{Background}
We review CeNNs and prior work with tracking algorithms.  

\subsection{Cellular Neural Networks}
\label{background:CNN}
\begin{figure}
    \begin{centering}
    \includegraphics[width=3.0in]{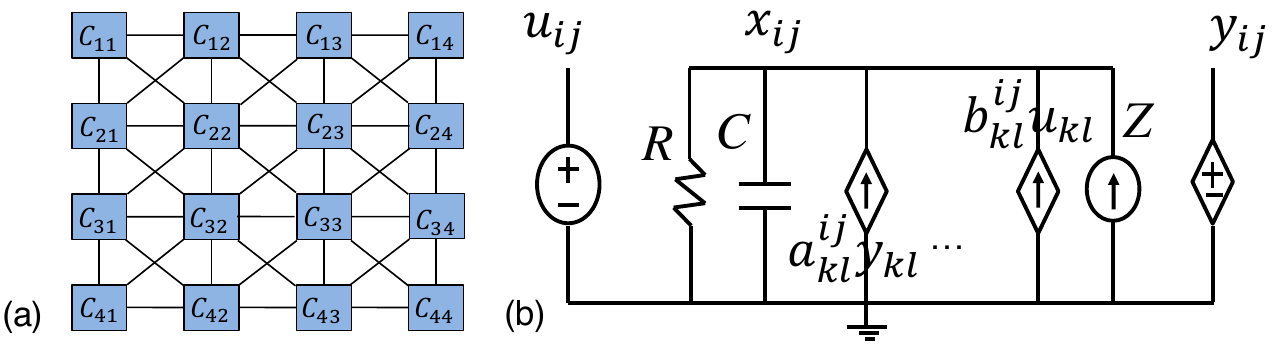}
    \caption{(a) CeNN architecture, (b) circuitry in CeNN cell.}
    \vspace{-0.1in}
    \label{fig:CNN}
    \end{centering}
 \end{figure}

A spatially invariant CeNN architecture \cite{CeNN} is an $M \times N$ array of identical cells (Fig. \ref{fig:CNN}a).  Each cell, $C_{ij}$, $(i,j)\in \{1, ..., M\} \times \{1, ..., N\}$, has identical connections with adjacent cells in a predefined neighborhood, $N_r(i,j)$ of radius $r$. The size of the neighborhood is $m=(2r+1)^2$, where $r$ is a positive integer. A conventional analog CeNN cell consists of one resistor, one capacitor, $2m$ linear VCCSs, one fixed current source, and one specific type of non-linear voltage controlled voltage source (Fig. \ref{fig:CNN}b). The input, state, and output of a given cell $C_{ij}$, correspond to the nodal voltages, $u_{ij}$, $x_{ij}$, and $y_{ij}$ respectively. VCCSs controlled by the input and output voltages of each neighbor deliver feedback and feedforward currents to a given cell. The dynamics of a CeNN are captured by a system of $M \times N$ ordinary differential equations, each of which is simply the Kirchhoff's Current Law (KCL) at the state nodes of the corresponding cells per  Eq.~\ref{eqn_conv_equilibrium}. 
\begin{align}
C\dfrac{dx_{ij}\left(t\right)}{dt} &= -\dfrac{x_{ij}\left(t\right)}{R}
+ \displaystyle\sum_{C_{kl} \in N_r\left(i,j\right)} a_{ij,kl}y_{kl}\left(t\right) \nonumber \\
&\qquad + \displaystyle\sum_{C_{kl} \in N_r\left(i,j\right)} b_{ij,kl}u_{kl} + Z
\label{eqn_conv_equilibrium}
\end{align}
CeNN cells typically employ a non-linear sigmoid-like transfer function \cite{Chua881} at the output to ensure fixed binary output levels. 

The parameters $a_{ij,kl}$, and $b_{ij,kl}$ serve as weights for the feedback and feedforward currents from cell $C_{kl}$ to cell $C_{ij}$. $a_{ij,kl}$, and $b_{ij,kl}$ are space invariant and are denoted by two $(2r+1) \times (2r+1)$ matrices.  (If $r=1$, they are captured by $3 \times 3$ matrices.)  The matrices of $a$ and $b$ parameters are typically referred to as the feedback template ($A$) and the feedforward template ($B$) respectively.  Design flexibility is further enhanced by the fixed bias current $Z$ that provides a means to adjust total current flowing into a cell. A CeNN can solve a wide range of image processing problems by carefully selecting the values of the $A$ and $B$ templates (as well as $Z$).

Various circuits including inverters, Gilbert multipliers, operational transconductance amplifiers (OTAs), etc \cite{Molinar07,Wang98} can be used to realize VCCSs. We use the OTA design from \cite{Qiuwen15}. OTAs provide a large linear range for voltage to current conversion, and can implement a wide range of transconductances allowing for different CeNN templates. 

\subsection{Target Tracking}
Given the initial position/size of a target object in a frame, a tracking algorithm should determine the position/size of the object in subsequent frames. Many studies have been performed to track objects in different scenarios \cite{Smeulders14,Hong15} (e.g., occlusion, illumination variation, etc. \cite{Wu13}).  Target tracking remains a challenging problem in computer vision as varying scene dynamics need to be reconciled in a single algorithm. 

Work in \cite{Timar05, rekeczky2002multi} initially considered CeNN-based trackers where CeNNs performed pixel-level computation, and a digital signal processor processed remaining information. However, \cite{Timar05, rekeczky2002multi} do not present a thorough energy/delay study. Moreover, \cite{Timar05, rekeczky2002multi} use a predefined sequence of kernels, and object tracking is based primarily on motion and change detection. While this may be useful if an object is moving continuously before a fixed background, these methods are not as general if used as a learning system.  As we will show in Sec. \ref{tracking}, our approach can select the optimal template sequences from a larger set in runtime based on the features of the tracked object, and the features of the background. 

When considering algorithms for von Neumann machines, convolution neural networks have been proposed to learn discriminative features or to help in the extraction of the discriminative saliency map \cite{Hong15,zhang2015robust}.  The outputs from a hidden layer can be used as feature descriptors.  The Struck algorithm \cite{Hare11} that we compare to is an adaptive tracking-by-detection methods for single target tasks. This is fundamentally an object recognition problem that places significant constraints on training/learning an object model. A tracker is initialized with a single bounding box provided as a ground truth for the starting frame of a video sequence. After initialization, the tracker maps each input frame to an output bounding box containing the predicted location of the object in that frame.  Struck uses a kernelized, structured, output support vector machine to provide adaptive tracking. It is well suited for drastic appearance changes for a tracked image in a given image sequence, and has the highest reported accuracy for the dataset that we use to evaluate our CeNN-based approach \cite{Wu13}.

To quantify accuracy, in this paper, we use a {\it success plot} to evaluate the bounding box overlap \cite{Wu13}. Given a tracked bounding box $r_t$ and the ground truth bounding box $r_a$, the success plot is defined as $S=\frac{|r_t \cap r_a|}{|r_t \cup r_a|}$, where $\cap$ and $\cup$ represent the intersection and union of two regions, respectively.


\section{Cellular neural network system}
Here, we discuss our CeNN-based co-processor and a framework for mapping a problem to such a co-processor.

\subsection{A CeNN-based co-processor}
\begin{figure}
    \centering
    \includegraphics[width=3.0in]{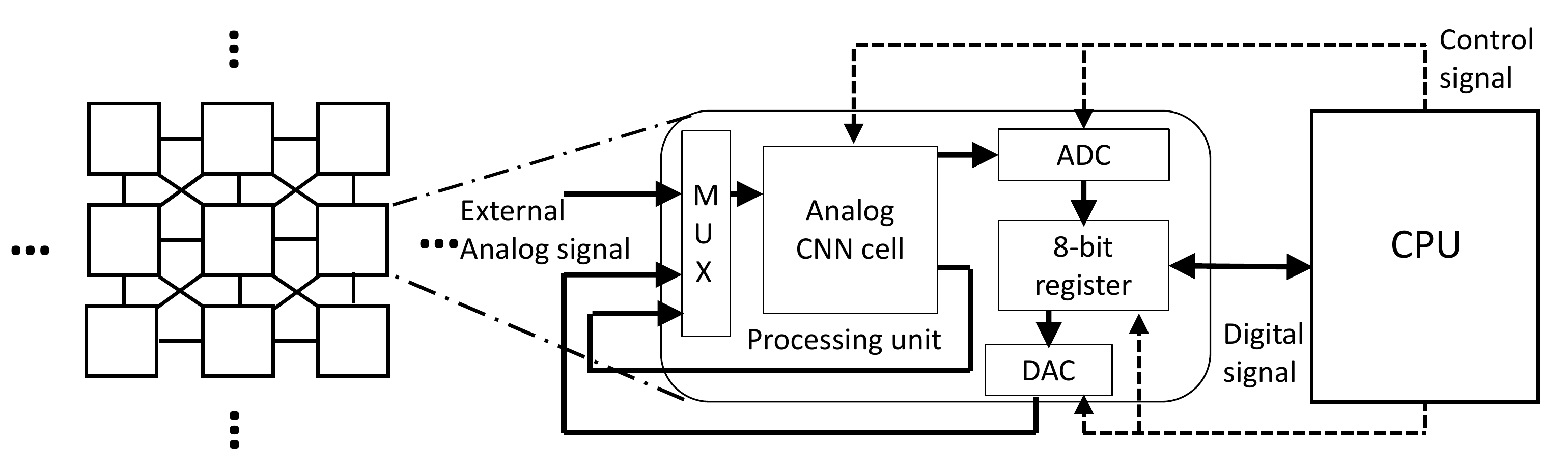}
    \caption{A CeNN based hardware accelerator.}
    \label{hw_acc}
    \vspace{-0.2in}
\end{figure}

The architecture and the processing unit schematics for our CeNN-based system are shown in Fig. \ref{hw_acc}.  Analog input signals (e.g., from sensors) are provided to the CeNN based co-processor. The CPU calls the co-processor to execute a given task/template by asserting a control signal. After processing, the CeNN based hardware accelerator generates an analog output which can either be converted to a N-bit digital signal and communicated with the CPU, or can be reused as an analog input to itself for further processing. Since each cell might need to send/receive data to/from the CPU, some mechanism is required to schedule the data movement. Each processing unit consists of a N-bit analog-to-digital converter, an analog CeNN cell, an analog multiplexer, and a N-bit register.  The analog multiplexer selects a signal either from the external environment, or the analog CeNN cell itself. The analog CeNN cell is as described in Sec. \ref{background:CNN}.

\subsection{A framework for evaluation}
\begin{figure}
    \centering
    \includegraphics[width=3.0in]{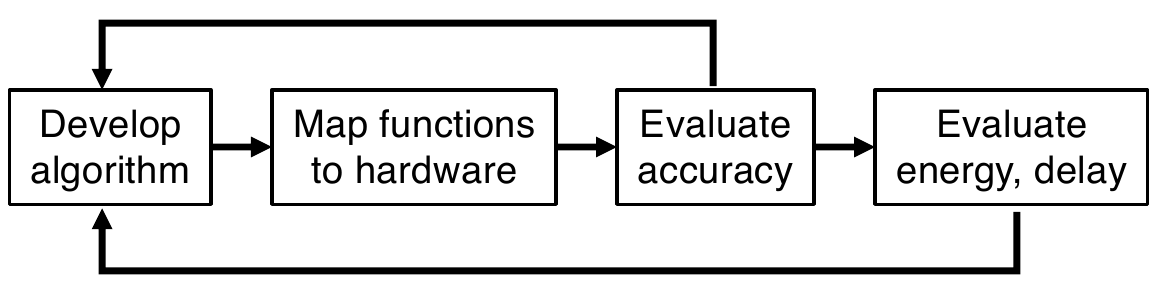}
    \caption{Framework for evaluating the CeNN hardware accelerator.}
    \label{frame_eval}
    \vspace{-0.2in}
\end{figure}

We now discuss an evaluation framework (see Fig. \ref{frame_eval}) to quantify the benefits of a CeNN co-processor. We must first develop an algorithm that: (i) is suitable for CeNN hardware, and (ii) achieves the same functionality as a traditional von-Neumann algorithm at application-level.  As CeNN arrays process information via template operations on local data, we first identify the global problem and describe it by local operations and descriptors. For target tracking, we must identify local features (e.g., orientation, movement direction, etc.), and the positions of detected features can be used to estimate the position of the target. For example, in the CeNN paradigm, the position of local features can be detected by {\it shadow templates} that can identify the vertical and horizontal bounding box sides of local features. (Per Sec. \ref{sec:4}, the local kernels used for feature extraction are selected via machine learning. A varying number of kernels can be selected which give a sparse and strong response on many images.)

After the algorithm is defined, its functional steps must be mapped on to suitable portions of the system. The algorithm is designed to maximize the use of CeNN hardware. However, portions of the algorithm might still require numerical operations on a CPU. Computation steps that involve parallel, independent processing of signals are mapped on to CeNN co-processors, e.g., feature extraction, classification at the pixel level, etc. Computations requiring precise floating point operations are mapped on to the CPU. The N-bit register within each processing unit of the CeNN architecture can facilitate any required data communication to and from the CPU.  \Note{Indranil: But this sounds like we get 8-bit precision from the analog CeNN.  Even if we extend to 32-bit, we still start with 8-bit.  Can you clarify this a bit?}

The accuracy of the algorithm's output must also be evaluated, and the algorithm may need to be refined to achieve suitable accuracy. Inaccuracies can be sourced from both the quality of the algorithm, and the analog nature of the hardware platform, and accuracy should be evaluated at multiple levels.

Finally, the algorithm and architecture must be evaluated to quantify the power, energy, and delays required for computation, and compared to the best von Neumann algorithm for the same problem. Analysis at this stage can lead to further modification of both the algorithm and architecture to maintain desired power-density of the co-processor, frame processing rates of the system (e.g., to enable real time tracking), etc.

\section{CeNN based target tracking algorithm}
\label{sec:4}
\label{tracking}

We now provide a detailed description of our CeNN-based target tracking algorithm.  The algorithm can be divided into two parts, training and tracking.   (While we evaluate the tracking portion -- i.e., to make an "apples-to-apples" comparison to Struck -- we also discuss the training portion for completeness.  Other approaches in the literature also consider image recognition with off-line training, and then report energy-delay-accuracy per classification --e.g., \cite{MNIST-2} for MNIST.)

\subsection{Training}
In the training process, we select a set of Difference of Gaussians (DoG) kernels from a feature pool, assign weight to each kernel, and select a threshold to profile the interest area. The selected parameters above can be used on subsequent images to identify the target.

Training is currently done off-line with a CPU. A set of features are selected that best match the target object and separates it from the features of the background and surrounding objects. The feature pool is generated by running a number of DoG kernels for features extraction on an image and pooling the results. A DoG kernel performs two isotropic or directional Gaussian diffusions along with a subtraction operation. 
Each DoG operation is followed by a pooling step, which also leverages fast-running CeNN templates. The most descriptive threshold is selected from a series of thresholds at different levels, and the binary result is diffused with an isotropic template. The result of this sequence of operations corresponds to different feature responses. This approach is similar to the operation used in deep learning.

From this image set, a subset of descriptor images can be selected to create a globally sparse and high-response area corresponding to the estimated object position, given as a bounding box. A user-specified number of features is selected from the pool based on response strength within the bounding box. The features are then weighted by a genetic algorithm. After the genetic algorithm produces optimal linear weights, the linear sum of the descriptors is thresholded at a value to highlight the interest area.

\subsection{Tracking}
\begin{figure}
    \centering
    \includegraphics[width=3.1in]{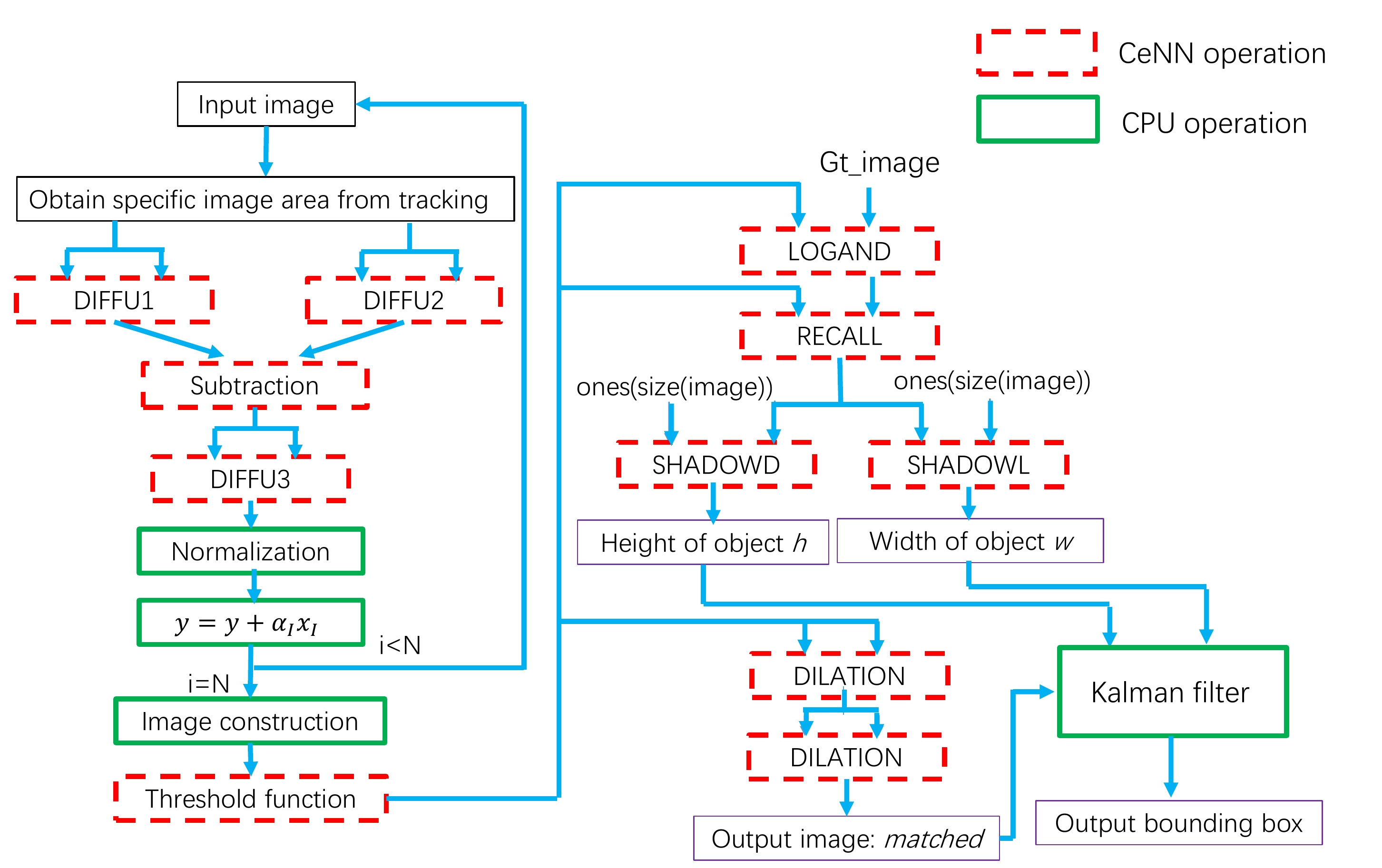}
    \caption{Processing flow for CeNN based tracking algorithm.}
    \label{tracking_algorithm}
    \vspace{-0.125in}
\end{figure}

Our CeNN-based tracking algorithm/information processing flow is shown in  Fig. \ref{tracking_algorithm}. (The remaining part is Kalman filter-based processing which is computed by the CPU.)
\Note{I updated the flow chart. I think we can eliminate the sentence in the bracket.}

The predefined DoG kernels, running by CeNN, are used in tracking to profile the object of interest. Each DoG kernel is weighted by a generic algorithm; the normalization and weighted sum of the image after DoG, performed by a CPU, would be the featured image. The runtime for each DoG operation is less than the settling time of a CeNN, so extra circuitry would be needed to control the timing of operation.  Then, tracking is performed by taking the centroid of the local response area of the current frame and resizing the bounding box according to the dimensions of the response area relative to the response area of the first frame.  After the image is constructed, several CeNN operations follow. A THRESHOLD operation can be performed to further highlight the object of interest. A following AND operation with the ground truth image can extract the object of interest. Once the extracted features are obtained, two SHADOW operations project the object vertically and horizontally. The width and height of the bounding box of the object can be easily calculated from said projections. A DILATION operation is used to augment the possible location of the tracked object, to ensure the object moves only a limited number of pixels between two consecutive frames. In the second part, the CPU applies two Kalman filters \cite{Lipton1998}. Weak Kalman filtering is also utilized to estimate the motion of the object, while a Strong Kalman filter is used to resize the bounding box, which is more prone to fluctuation.  Once the position estimation is known the second part of the algorithm is executed on the next frames iteratively. 



\section{Target tracking alogrithm evaluation}

We now present our simulation/measurement setup and energy/delay/accuracy projections for the two algorithms.

\subsection{Image sequences to benchmark}
To quantitatively evaluate energy, delay, and accuracy for both algorithms, we selected 24 video sequences from \cite{benchmark}. These test sequences are categorized into three sets (with 8 sequences per set).  Dataset 1 consists of image sequences where we expect good overlap per a success plot. These sequences exhibit the following characteristics:  (i) the motion of the object to be tracked is either in one direction or very slow, (ii) the object is easy to identify from the background, and (iii) the shape of the object does not change significantly.  In Dataset 2, typically one of the three aforementioned characteristics of scene dynamics is {\it not} satisfied.  Finally, in Dataset 3, most/all of the aforementioned conditions are not satisfied. 

\subsection{Simulation setup}
\label{evaluation:setup}
Some portions of our CeNN-based algorithm are run on a CPU, while all code for the Struck algorithm is run on a CPU.  We use source code for Struck from \cite{Struck}. Any CPU code needed for our CeNN approach was written in C.  All CPU code was then executed on an Intel Core i5-5675C, 16GB DDR3, with Ubuntu 15.04 64-bit. We use a setup like that described in \cite{tang2013gpu} to measure the energy usage of the CPU. The CPU is powered by a dedicated 12V cable. We use a current clamp to monitor the current changes on the cable at a speed of 10,000 samples per second. A data acquisition system records the current data when the CPU executes code for a respective algorithm.  
We start the timer after the CPU reads the image, and synchronize the execution time. Thus, we only compute the delay/energy of the CPU. Memory overhead is not considered as most program data is stored in registers for both architectures. For the CeNN based algorithm, we also set the timer at the beginning and end of CPU operations to eliminate other overheads. We then use current, voltage and time data to compute CPU energy.

\subsection{Delay and energy analysis}
\label{evaluation:energy}
Here, we present energy/delay data for image sequences in Dataset 1.  (Metrics for other datasets were also collected, but projected gains in EDP were similar to Dataset 1.)
For the CeNN algorithm, we report data for the CPU as well as the CeNN hardware accelerator.  
For the hardware accelerator, we assume 14 nm CMOS, the same technology node as the CPU.

Per Fig. \ref{tracking_algorithm}, a portion of the tracking algorithm is mapped to the CeNN co-processor, while the rest is mapped to the CPU.  We first consider the steps that leverage the CeNN accelerator, and summarize energy and delay in Table \ref{tbl:cnn_steps}, (using the {\it RedTeam} image sequence from Dataset 1 as an example). Time and energy is dominated by some image augmentation and shadow steps (e.g., RECALL and SHADOWL). As these are propagating type of templates, the run-time of these steps need to be long enough for signals to propagate through the whole image.
The initialization reflects the data movement from other components to the hardware accelerator. In this process, the state voltage should be pre-charged before running the CeNN operation.  In the initialization step, the data either comes from the CPU through a DAC, or from a photo sensor.  The run-time for the initialization includes the the settling time for the CeNN cell, and the run-time of the DAC (if data comes from a CPU). 

The settling time for a single cell is computed by \cite{iccad-14}. The power for OTAs (which can dominate CeNN cells and is computed per \cite{iccad-14}) is constant in operation, and proportional to the $G_m$ of the OTA \cite{iccad-14}. $G_m$ is proportional to the template value, which is specified in \cite{template_library}. We designed an OTA for a specific template value, (i.e., template values are 1), measured its power via SPICE, and scaled the power based on numerical values of given template values. In this way, we obtain the power of all OTAs in a CeNN cell for one operation and ultimately the power for the analog CeNN co-processor. We iteratively follow the previous step to obtain the power for each step, and use the measured delay to obtain energy data.


Note that we assume the run-time of the DAC is the same as ADC \cite{ADC14}. 
The energy for the ADC is calculated per \cite{ADC14}. 
The frequency of the ADC is determined by the number of ADC accesses dividing the total run-time of analog CeNN component.  The frequency of the register is the same as the frequency of the ADC to ensure data synchronization. The bit length for the ADC, register, and DAC is assumed to be 8 which provides enough ($256$) shades of grays for image processing.



\begin{table}
 \caption{Time and energy for CeNN in each step}
 \centering
 \scriptsize
 \newcommand{\tabincell}[2]{\begin{tabular}{@{}#1@{}}#2\end{tabular}}
\renewcommand{\arraystretch}{1.35}
 \begin{tabular}{| c |c |c | c |c| }
 \hline
  Operations &  \tabincell{c}{Time/frame\\($\mu$s)}  & \tabincell{c}{OTA power\\($\mu$W)}  & \tabincell{c}{Total power\\($\mu$W)} & \tabincell{c}{Energy per\\ frame ($\mu$J)} \\
 \hline
  Init & $7.5$ & $1.5$ & $1.58$ & $0.0081$ \\
 \hline
    DIFFUS1 & $7.5$ & $1.5$ & $1.58$ & $0.0081$ \\
 \hline
    Init & $7.5$ & $1.5$ & $1.58$ & $0.0081$ \\
 \hline
    DIFFUS2 & $7.5$ & $1.5$ & $1.58$ & $0.0081$ \\
 \hline
    SUB & $7.5$ & $3.0$ & $3.08$ & $0.0158$ \\
 \hline
    DIFFUS3 & $25$ & $1.5$ & $1.58$ & $0.027$ \\
 \hline
    Init & $0.3$ & $1.5$ & $1.58$ & $0.0324$ \\
 \hline
    THRES & $0.3$ & $3.0$ & $3.08$ & $0.0781$ \\
 \hline
    LOGAND & $0.3$ & $6.0$ & $6.08$ & $0.154$ \\
 \hline
    RECALL & $35.2$ & $21.8$ & $21.8$ & $64.9$ \\
 \hline
    SHADOWL & $35.2$ & $9.0$ & $9.08$ & $27$ \\
 \hline
    Init & $0.3$ & $1.50$ & $1.58$ & $0.00324$ \\
 \hline
    SHADOWD & $24$ & $9.0$ & $1.58$ & $18.4$ \\
     \hline
    DILATION & $1.6$ & $13.5$ & $13.6$ & $1.84$ \\
     \hline
    Total/frame & $160$ &  & $76.9$ & $112$ \\
     \hline
    Total (all frames) & $0.306$s &  &  & $0.216J$ \\
  \hline
  \end{tabular}
  \label{tbl:cnn_steps}
  \vspace{-0.1in}
\end{table}

\begin{figure}
    \centering
    \includegraphics[width=3.5in]{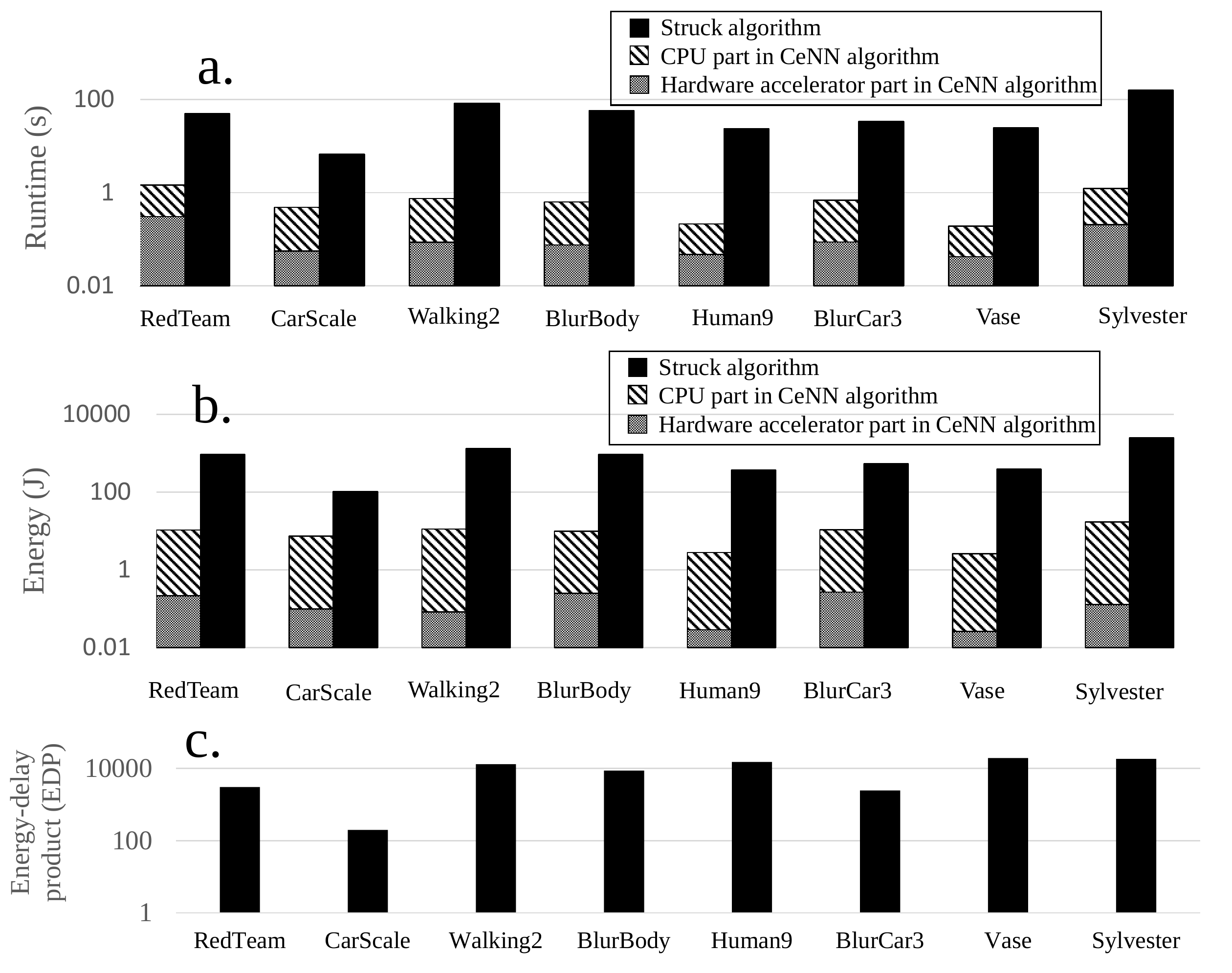}
    \caption{(a) Run time and (b) energy for the CeNN based algorithm and Struck algorithm. The stacked bars are for the CeNN algorithm -- dotted bar for the CeNN part, bar with diagonal lines for the CPU part (c) energy-delay product (EDP) of Struck algorithm over the CeNN algorithm.}
    \label{fig:edp}
    \vspace{-0.1in}
\end{figure}

Delay, energy, and EDP for the CeNN-based system is depicted by the stacked bars in Fig.~\ref{fig:edp}(a)-(c) respectively. (The dotted bars represent the CeNN-based co-processor, while the bars with diagonal lines represent computations mapped to the CPU.) The rightmost solid bar represents the Struck algorithm. Here, computations mapped to the CPU in the CeNN based algorithm generally have higher delay and energy when compared to information processing operations performed on the CeNN.  

To fully appreciate the benefit of CeNNs, we have run a version of the CeNN-based algorithm implemented in Matlab on a CPU. Run times are on the order of a few hours for a given image sequence.  Given that the corresponding C code for CeNN algorithm 
runs 100X faster, the Matlab-based approach is still orders of magnitude slower than the system containing a CeNN accelerator. Furthermore, our software implementation indicates that 99\% of the computation time is attributed to CeNN operations. Hence, the CeNN co-processor has a substantial, positive impact on the CeNN-based system.

Finally, for the Struck algorithm, energies and delays for various sequences are directly obtained from measured CPU data per Sec. \ref{evaluation:setup}.  For the different image sequences in Fig.~\ref{fig:edp}, EDPs are largely 1000X to 10000X better for our CeNN-based approach versus Struck.



\subsection{Accuracy analysis}
\begin{figure}
    \centering
    \includegraphics[width=3.0in]{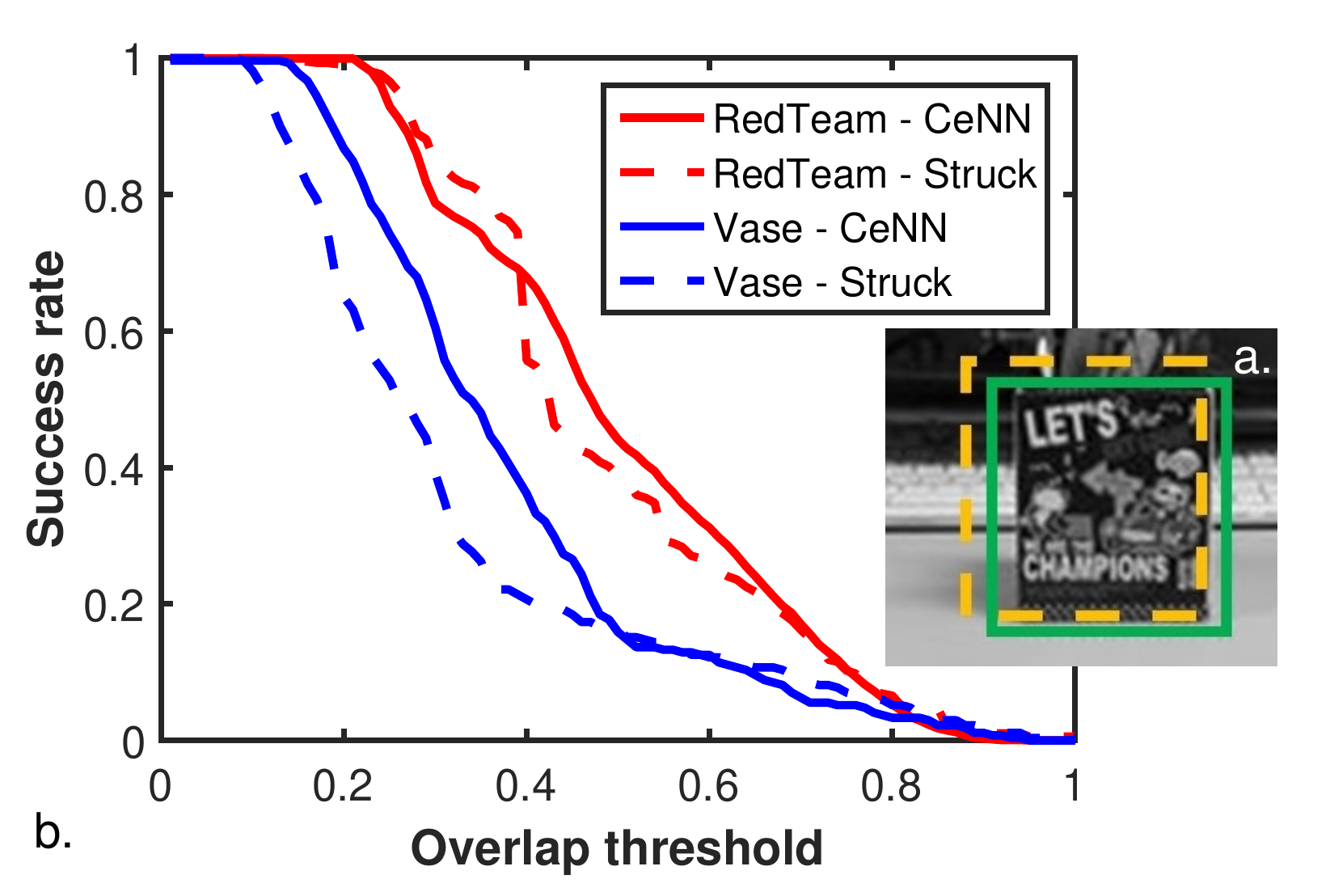}
    \caption{Accuracy for RedTeam and Vase benchmarks.}
    \label{AUC}
    \vspace{-0.1in}
\end{figure}

Given the potential for energy/delay improvements, we now consider tracker accuracy.  In Fig. \ref{AUC}, we show representative results for accuracy for the {\it RedTeam} and {\it Vase} image sequences from Dataset 1.  Per the inset, the solid box represents the ground truth, while the dashed box represents the position established by the tracker.  The {\it overlap threshold} is the intersection of the two boxes over the union of the two boxes.  After collecting this data for every image in the sequence, we can plot success rate (i.e., the percentage of time a given overlap threshold was achieved).  For a perfect tracker, this plot would simply be a horizontal line at '1' on the y-axis.  To represent accuracy as a single number, we calculate the area under the curve (AUC) for a given image sequence and algorithm.

AUC scores were calculated for the 24 sequences for both Struck and our CeNN-based approach.  For Struck, the algorithm can be tuned such that energy, accuracy, or run-time may vary (e.g., leveraging more kernels, changing the search radius etc.). We set these parameters to obtain the highest possible accuracy as per Sec. \ref{evaluation:energy}, our CeNN-based approach can be superior to a von Neumann approach in terms of energy and delay.  As such, we consider these metrics as we approach iso-accuracy.  To obtain AUC data for our CeNN-based approach, we used a Matlab-based CeNN simulator \cite{horvath14}. While both the Matlab simulator and Struck algorithm run on a 64-bit machine, the image quality is 8-bits. Thus, CeNN hardware should be sufficient to process these video sequences.  \Note{Could someone argue that Struck could be better too if run on an 8-bit machine?}


\begin{figure}
    \centering
    \includegraphics[width=3.25in]{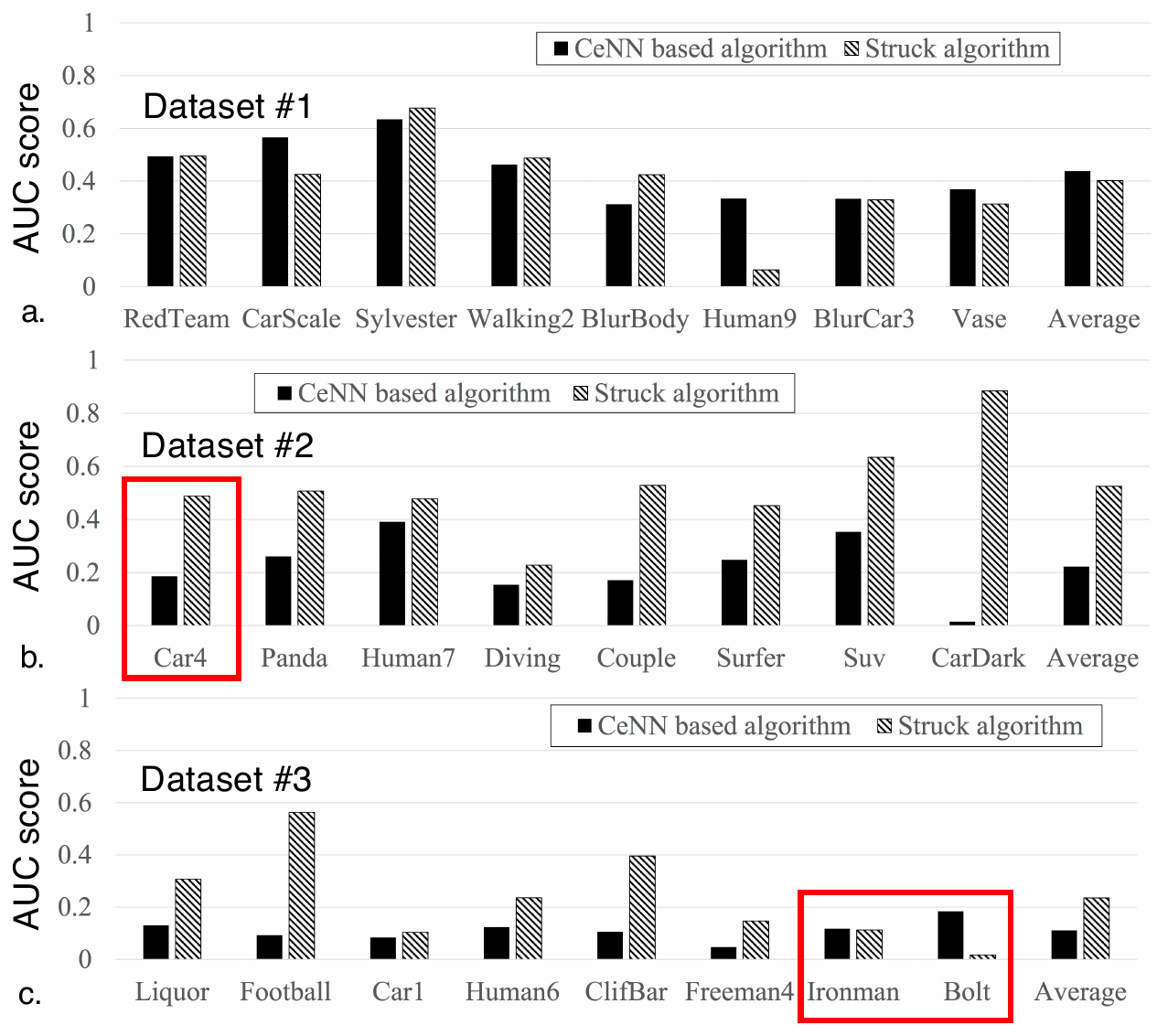}
    \caption{AUC score for (a) Dataset 1, (b) Dataset 2, and (c) Dataset 3.}
    \label{accuracy}
    \vspace{-0.2in}
\end{figure}


The accuracy results (AUC scores for all 24 benchmarks) are shown in Fig. \ref{accuracy}.  For dataset 1, the average AUC for our algorithm and Struck are 0.438 and 0.402, respectively (see Fig. \ref{accuracy}a). As scene dynamics change such that an object becomes more difficult to track, the AUC scores for both algorithms decrease.  However, for Datasets 2 and 3 (Figs. \ref{accuracy}b, c) the Struck-based approach is approximately 2X better than the CeNN algorithm for the AUC benchmark.  That said, there are exceptions.  Among the most challenging image sequences (i.e., Ironman and Bolt), the CeNN approach is either similar or superior to Struck (see highlighted sequences in Fig. \ref{accuracy}c).
While we will consider this more at the algorithm-level in future work, we briefly discuss our initial hypothesis as to why this is the case.  Using the Bolt dataset as an example -- where there are several athletes running -- the CeNN algorithm can focus more on local features with an object moving only a limited number of pixels in two consecutive frames.



Additionally, accuracy projections for the CeNN-based algorithm currently only consider basic configurations of the CeNN-based algorithm, which can be further improved.  As examples, the number of kernels can be increased to capture more detailed 
features.  CeNN connectivity can also be increased to expand local feature area. Gabor kernels will increase sensitivity to spatial frequencies and the orientation of edges and enhance feature extraction. To quantitatively address this hypothesis, we considered the Car4 benchmark (see highlighted sequence in Fig. \ref{accuracy}b) from Dataset 2 with initial AUC score of 0.186. When we increase the number of kernels from 25 to 50, and CeNN connectivities from $3\times 3$ to $5\times 5$, AUC increases to 0.200. If we employ CPU-based Gabor kernels the AUC score improves to 0.378.  While this will likely increase energy and delay, per Sec. \ref{evaluation:energy}, a substantial margin exists.  (Gabor kernels can also be efficiently realized with CeNNs \cite{Gabor-CNN}).  This will be studied in future work.

\vspace{-0.08in}
\section{Conclusion}
We have presented a framework for evaluating CeNN-based co-processors.  As a case study, we developed a CeNN-based target tracking algorithm for a CeNN-based co-processor, and compared its energy/delay/accuracy to a highly accurate, von Neumann tracker.  Our CeNN-based tracker is as accurate as the Struck algorithm for many image sequences, and can offer orders of magnitude improvements in EDP.  When Struck is superior, we have outlined paths to algorithmic improvement that makes our approach competitive.  In future work, we will (i) accelerate the training process with CeNNs, (ii)  evaluate energy/delay with algorithmic refinements, and (iii) study the impact of other emerging technologies (e.g., which may enable highly efficient/compact diffusion networks \cite{behnamSymFET}).  


\bibliographystyle{IEEEtran}
\bibliography{CNNcitations}

\end{document}